\documentclass[aps,twocolumn,epsfig,graphics,showpacs,floatfix,mathbbm]{revtex4}

\usepackage{amsmath,amsfonts,amssymb,graphics,graphicx,epsfig,color}
\usepackage{epsfig}
\usepackage{amscd}
\usepackage{latexsym}
\usepackage{graphicx}

\newtheorem{lem}{Observation}
\newtheorem{theo}{Theorem}

\newcommand{\ket}[1]{\ensuremath{|#1\rangle}}

\newcommand{\ketbra}[1]{\ensuremath{| #1 \rangle \langle #1 |}}

\newcommand{\fA}{\ensuremath{\mathfrak{A}}}

\newcommand{\BE}{\begin{equation}}
\newcommand{\EE}{\end{equation}}
\newcommand{\be}{\begin{equation}}
\newcommand{\ee}{\end{equation}}
\newcommand{\bea}{\begin{eqnarray}}
\newcommand{\eea}{\end{eqnarray}}
\newcommand{\kommentar}[1]{}

\begin{document}

\title{Detecting two-party quantum correlations in quantum key
distribution protocols}
\author{Marcos Curty$^1$, Otfried G\"{u}hne$^{2,3}$, Maciej Lewenstein$^2$, and Norbert L\"{u}tkenhaus$^1$}
\affiliation{$^1$Quantum Information Theory Group, Institut f\"ur
Theoretische Physik I, and Max-Planck Research Group, Institute
of Optics, Information and Photonics, Universit\"{a}t
Erlangen-N\"{u}rnberg, Staudtstra{\ss}e 7/B2, 91058
Erlangen, Germany \\
$^2$Institut f\"{u}r Theoretische Physik, Universit\"{a}t
Hannover, Appelstra{\ss}e 2, 30167 Hannover, Germany \\
$^3$Institut f\"{u}r Quantenoptik und Quanteninformation,
\"Osterreichische Akademie der Wissenschaften, 6020 Innsbruck,
Austria}

\begin{abstract}
A necessary precondition for secure quantum key distribution
(QKD) is that sender and receiver can prove the presence of
entanglement in a quantum state that is effectively distributed
between them. In order to deliver this entanglement proof one can
use the class of entanglement witness (EW) operators that can be
constructed from the available measurements results. This class
of EWs can be used to provide a necessary and sufficient
condition for the existence of quantum correlations even when a
quantum state cannot be completely reconstructed. The set of
optimal EWs for two well-known {\it entanglement based} (EB)
schemes, the $6$-state and the $4$-state EB protocols, has been
obtained recently [M. Curty et al., Phys. Rev. Lett. {\bf 92},
217903 (2004)]. Here we complete these results, now showing
specifically the analysis for the case of {\em prepare\&measure}
(P\&M) schemes. For this, we investigate the signal states and
detection methods of the $4$-state and the $2$-state P\&M
schemes. For each of these protocols we obtain a reduced set of
EWs. More importantly, each set of EWs can be used to derive a
necessary and sufficient condition to prove that quantum
correlations are present in these protocols.
\end{abstract}

\maketitle

\section{INTRODUCTION}

One of the most important problems in modern cryptography is the
transmission of secret information from a sender (usually called
Alice) to a receiver (Bob) over an insecure communication channel
\cite{SCHNEIER_1996}. The goal is to guarantee that any possible
eavesdropper (Eve), with access to the channel, is unable to
obtain useful information about the message.

Secret systems were studied from an information-theoretic
perspective by C.~E. Shannon \cite{Shannon49}. He analyzed the
natural scenario where Eve has always access to exactly the same
information received by Bob. This information, denoted as $C$
(from the term ciphertext), is typically obtained by Alice as a
function of the message to be sent, $M$, and a secret key, $K$,
that she needs to share previously with Bob. In this context,
Shannon defined a cryptographic system to be perfectly-secret and
uniquely decodable if it satisfies the following two conditions:
first, the ciphertext, $C$, and the message, $M$, must be
statistically independent. This means that Eve cannot obtain any
useful information about the message $M$ from $C$. This condition
can be expressed as $I(M;C)=0$, where $I$ denotes the mutual
information measured in bits \cite{Shannon49}. The second
condition states that Bob can recover the original message $M$
from $C$ and $K$. It can be formulated as $H(M|C,K)=0$, with $H$
the Shannon entropy measured also in bits \cite{Shannon49}. With
this definition, Shannon proved the well-known pessimistic result
that every perfectly-secret uniquely decodable system must
satisfy $H(K)\geq{}H(M)$. An example of a secret cryptographic
system satisfying this condition is the so-called one-time-pad or
Vernam cipher \cite{Vernam26}.

The result from Shannon relies in a fundamental way on considering
that both Bob and Eve have perfect access to the same ciphertext
$C$. However, there are scenarios, such is the case in quantum key
distribution (QKD), where the proper exploitation of particular
quantum effects can prevent Bob and Eve to receive precisely the
same information. The laws of Quantum Mechanics can guarantee some
minimal uncertainty between both signals, and this fact can be
used by Alice and Bob to expand a previously shared secret key
$K$ in an unconditionally secure manner
\cite{Aut,BB84,Ekert91,Bruss98}. This means that QKD together
with the Vernam cipher can in principle be used to achieve
perfectly secret communications even when $H(K)\ll{}H(M)$.

In any realization of QKD one can typically distinguish two
phases in order to expand a secret key. In the first phase, an
effective bi-partite quantum mechanical state is distributed
between Alice and Bob. This state creates correlations between
them and it might contain as well hidden correlations with Eve.
Next, a (restricted) set of measurements is used by the
legitimate users to measure these correlations. As a result,
Alice and Bob obtain a classical joint probability distribution
$P(A,B)$ representing the measurement results. In the second
phase, usually called {\em key distillation}, Alice and Bob use
an authenticated public channel to process the correlated data in
order to obtain a secret key. This procedure involves, typically,
postselection of data, error correction to reconcile the data,
and privacy amplification to decouple the data from Eve
\cite{Norbert99}.

Two types of schemes are used to create the correlated data in
the first phase of QKD. In {\em entanglement based} (EB) schemes
an, in general, untrusted third party distributes a bi-partite
state to Alice and Bob. This party may be even Eve who is in
possession of a third sub-system that may be entangled with those
given to Alice and Bob. While the subsystems measured by Alice
and Bob result in correlations described by $P(A,B)$, Eve can use
her subsystem to obtain information about the data of the
legitimate users.

In {\em prepare\&measure} (P\&M) schemes Alice prepares a random
sequence of pre-defined non-orthogonal states $|\varphi_i\rangle$
that are sent to Bob through an untrusted quantum channel
(possibly controlled by Eve). On the receiving side, Bob performs
a positive operator value measure (POVM) on every signal he
receives. Generalizing the ideas introduced by Bennett {\it et
al.} \cite{Mermin92}, the signal preparation process in P\&M
schemes can be thought of as follows: Alice prepares an entangled
bi-partite state of the form $|\Psi\rangle_{AB} =\sum_i \sqrt{p_i}
|\alpha_i\rangle|\varphi_i\rangle$, where the states
$|\alpha_i\rangle$ form an orthonormal basis and $\{p_i\}_i$
represents the {\it a priori} probability distribution of the
signal states $|\varphi_i\rangle$. If now Alice measures the
first system in the basis $|\alpha_i\rangle$, she effectively
prepares the (non-orthogonal) signal states $|\varphi_i\rangle$
with probabilities $p_i$. The action of the quantum channel on the
state $|\Psi\rangle_{AB}$ leads to an effective bi-partite
quantum state shared by Alice and Bob. One important difference
between P\&M schemes with effective entanglement and EB schemes
with real entanglement is that in the first case the reduced
density matrix of Alice,
$\rho_A=Tr_B(|\Psi\rangle\langle\Psi|_{AB})$, is fixed and known
and cannot be modified by Eve.

An essential question in QKD now is whether the correlated data
generated in the first phase enable Alice and Bob to extract a
secret key. In Ref.~\cite{Curty04} it has been proven that a
necessary precondition for secure key distillation is the
provable presence of quantum correlations in $P(A,B)$. That is,
it must be possible to interprete $P(A,B)$, together with the
knowledge of the corresponding measurements, as coming {\it
exclusively} from an entangled state. Moreover, this result
applies both for EB and P\&M schemes (For EB schemes see also
\cite{GisinWolf00}). Alice and Bob need to be able to detect the
presence of entanglement in the quantum state that is effectively
distributed between them, otherwise no secret key can be
obtained. Among all separability criteria available nowadays to
deliver this entanglement proof (see, e.g., \cite{Separability}
and references therein), entanglement witnesses (EWs)
\cite{Horodecki96,Terhal00,Lewenstein00} are particularly suited
for this purpose since they give rise to a necessary and
sufficient condition for the existence of quantum correlations in
$P(A,B)$, even when the state shared by Alice and Bob cannot be
completely reconstructed \cite{Curty04}. In Ref.~\cite{Curty04} a
detailed analysis of two well-known EB protocols, the $6$-state
and the $4$-state EB protocols \cite{Bruss98,BB84,Ekert91}, is
included and the set of optimal EWs to detect quantum
correlations in both protocols has been found. The purpose of
this paper is to complete the results contained in
Ref.~\cite{Curty04}, now showing specifically the analysis for
the case of P\&M schemes. In particular, we investigate the
signal states and detection methods of the $4$-state and the
$2$-state P\&M schemes \cite{BB84,Ben92}, and we obtain a reduced
set of EWs that can be used to derive a necessary and sufficient
condition to prove that quantum correlations are present in these
protocols. As a side point, we put into context recent results
that can be useful in the search of quantum correlations for
higher dimensional QKD schemes \cite{jens04}.

The paper is organized as follows. In Section \ref{QC} we review
the role of quantum correlations as precondition for secure QKD.
Section \ref{EW} introduces the concept of EWs and shows how to
detect quantum correlations by using the class of EW operators
that can be constructed from the available data. This formalism
is then used in Section \ref{PROT} to analyze well-known QKD
protocols. Our starting point are the EB schemes studied in
Ref.~\cite{Curty04}: The $6$-state and $4$-state EB schemes. Then
we present the new results for P\&M schemes, analyzing in detail
the $4$-state and the $2$-state P\&M schemes. The last part of
the section gives a brief outlook to the study of quantum
correlations in higher dimensionsal QKD schemes and in practical
QKD. Finally, Section \ref{Concl} concludes the paper with a
summary.

\section{QUANTUM CORRELATIONS \& QUANTUM KEY DISTRIBUTION}
\label{QC}

As mentioned in the introduction, the provable presence of
quantum correlations in $P(A,B)$ has been shown to be a necessary
precondition for secure QKD \cite{Curty04}. The starting point for
such a proof is an upper bound for the distillation rate of a
secure key from correlated data via authenticated public
communication, which is given by the {\em intrinsic information }
$I(A;B\downarrow{}E)$, introduced by Maurer and Wolf
\cite{Maurer99}. These authors considered the problem of key
distillation in the classical scenario where Alice, Bob, and Eve
have access to repeated independent realizations of three random
variables, denoted as $A$, $B$, and $E$, characterized by a
probability distribution $P(A,B,E)$. In this context, Maurer and
Wolf proved that the rate of secret bits, denoted as $S(A;B||E)$,
that Alice and Bob can get by communicating to each other through
a public authenticated channel satisfies \cite{Maurer99}:

\begin{equation}
S(A;B||E)\leq{}I(A;B\downarrow{}E) = \min_{E \to \bar{E}}
I(A;B|\bar{E}),
\end{equation}

\noindent where the minimization runs over all possible channels
$E \to \bar{E}$ characterized by the conditional probability
$P(\bar{E}|E)$, and $I(A;B|\bar{E})$ is the mutual information
between Alice and Bob given the public announcement of Eve's data
based on the probabilities $P(A,B,\bar{E})$. This quantity is
defined in terms of the conditional Shannon entropy
$H(X|\bar{e})=\sum_{x\in X} - p(x|\bar{e}) \log_2 p(x|\bar{e})$ as

\begin{equation}
I(A;B|\bar{E}) = \sum_{\bar{e}\in \bar{E}} P(\bar{e})
\Big[H(A|\bar{e})+H(B|\bar{e})-H(A,B|\bar{e})\Big].
\end{equation}

More important for QKD, the result of Maurer and Wolf can as well
be adapted to the case where Alice, Bob and Eve start sharing a
tri-partite quantum state instead of a joint probability
distribution. For this purpose, one can consider all possible
tri-partite states that Eve can establish using her eavesdropping
method, and all possible measurements she could perform on her
sub-system. This gives rise to a set of possible extensions
${\cal P}$ of the observable probability distribution $P(A,B)$ to
$P(A,B,E)$. Now one can define the intrinsic information as
\begin{equation}
I(A;B\downarrow{}E) = {\rm inf}_{{\cal P}}\;I(A;B|E) \; .
\end{equation}

The main consequence of this fact is that whenever the observable
data $P(A,B)$ can be explained as coming from a tri-partite state
with a separable reduced density matrix for Alice and Bob, the
intrinsic information vanishes and therefore no secret key can be
established.
\begin{lem}
\cite{Curty04} Assume that the observable joint probability
distribution $P(A,B)$ together with the knowledge of the
corresponding measurements performed by Alice and Bob can be
interpreted as coming from a separable state $\sigma_{AB}$. Then
the intrinsic information vanishes and no secret key can be
distilled via public communication from the correlated data.
\end{lem}

{\it Proof:} This is easy to see for EB schemes  as we extend a
separable reduced density matrix $\sigma_{AB}= \sum_i q_i
|\phi_i\rangle_A\langle \phi_i|\otimes
|\psi_i\rangle_B\langle\psi_i|$ to a tri-partite pure state of the
form $|\Phi\rangle_{ABE}= \sum_i \sqrt{q_i} |\phi_i\rangle_A
|\psi_i\rangle_B|e_i\rangle_E$. (See also \cite{GisinWolf00}.)
Here $|e_i\rangle_E$ is a set of orthonormal vectors spanning a
Hilbert space of sufficient dimension. If Eve measures her
sub-system in the corresponding basis, the conditional
probability distribution conditioned on her measurement result
factorizes such that for this measurement $I(A;B|E)=0$. As a
consequence, the intrinsic information vanishes and no secret key
can be distilled.

In the case of P\&M schemes we need to show additionally that the
state $|\Phi\rangle_{ABE}$ can be obtained by Eve by interaction
with Bob's system only. The initial state $|\Psi\rangle_{AB}
=\sum_i \sqrt{p_i} |\alpha_i\rangle|\varphi_i\rangle$ can be
written in the Schmidt decomposition as $|\Psi\rangle_{AB} =
\sum_i c_i |u_i\rangle_A |v_i\rangle_B$. Then the state
$|\Phi\rangle_{ABE}$ from above is in the Schmidt decomposition,
with respect to system $A$ and the composite system $BE$, of the
form $|\Phi\rangle_{ABE}= \sum_i\ c_i |u_i\rangle_A
|\tilde{e}_i\rangle_{BE}$ since $c_i$ and $|u_i\rangle_A$ are
fixed by the known reduced density matrix
$\rho_A=Tr_B(|\Psi\rangle\langle\Psi|_{AB})$ to the corresponding
values of $|\Psi\rangle_{AB}$. Then one can find always a
suitable unitary operator $U_{BE}$ such that
$|\tilde{e}_i\rangle_{BE}=U_{BE}|v_i\rangle_B|0\rangle_E$ where
$|0\rangle_E$ is an initial state of an auxiliary system.
$\blacksquare$

The natural question that arises now is whether the presence of
quantum correlations is also a sufficient condition for secure
QKD. Let us mention already here that this is still an open
question in the field of quantum cryptography. In EB schemes, it
is clear that it is possible to obtain a secret key whenever the
distributed bi-partite states are entangled qubit states {\em
and} each party is allowed to perform collective quantum
manipulations on their respective states. This is true since in
this situation one can first distill maximally entangled states
from the initial states and subsequently measure them out in the
standard basis \cite{Horodecki97}. The verification that the
entanglement distillation process succeeded allows to give the
security statement about the resulting perfectly correlated and
random measurement data, which can then be used as a secret key.

A completely different scenario arises once Alice and Bob have
already performed their respective measurements on the given
states and they can only use classical operations on their
correlated data. This last case has been partially addressed
under additional assumptions, namely that the eavesdropping attack
employed by Eve is restricted to the so-called ``incoherent
symmetric strategies'', in \cite{Gisin99}. In this situation it
has been proven that for a particular class of QKD protocols key
distillation is possible if and only if the initially distributed
states are distillable \cite{Gisin99}. In the same spirit,
Ac\'{\i}n et al. \cite{Acin03b} showed that one can always
distill a secret key from any two-qubit and one-copy distillable
states by adapting the local measurements to the quantum states
and performing subsequently a classical protocol. All these
results suggested the idea of a correspondence between
entanglement distillation and secret key distillation. See also
\cite{Winter1}. The main conjecture was that a quantum state
could lead to a secret key if and only if it is distillable,
which is not equivalent to containing quantum correlations
\cite{Phoro97}. However, this point of view changed recently,
since it has been shown that it is also possible to generate a
secret key even from certain non-distillable entangled states,
known as bound entangled or positive partial transposed (PPT)
entangled states \cite{karol}. These are states that require
entanglement to be created but do not allow to distill
entanglement from them \cite{Phoro97}. This shows that the focus
on entanglement distillation guided protocols in QKD is too
narrow, though the interesting example introduced in
Ref.~\cite{karol} does not answer the question whether all
entangled states can be transformed into a private key.

More recently, going back to the quantum correlations point of
view, Ac\'{\i}n and Gisin \cite{Acin03aSub} proved that it is an
equivalent statement to show that there has been (real or
effective) entanglement in the distributed quantum state and that
the intrinsic information is non-zero. In particular, this result
implies that there exist a one-to-one relation between the
detection of entanglement in $P(A,B)$ and the fact that such
probability distribution cannot be obtained by classical means
using only local operations and classical communication
\cite{Acin03aSub,renn}. That is, $P(A,B)$ contains secret bits.
More important for QKD, this means that either it is possible to
distill a secret key from {\it any} bi-partite entangled state or
there exits a classical analog of bound entanglement, the
so-called bound information \cite{gisinwolf}. This is information
shared by Alice, Bob and Eve such as Alice and Bob cannot obtain
a secret key from it although this information cannot be
distributed by local operations and classical communication.
However, so far the existence of bound information has been
proven for the multi-partite case \cite{cirac} (for the case of
coherent manipulations of multi-party quantum states see also
\cite{augu}), but not for the bi-partite case relevant for QKD.

\section{DETECTING QUANTUM CORRELATIONS} \label{EW}
Given that quantum correlations are necessary for distilling a
secure secret key, the question now is how to detect these
quantum correlations in a given QKD scheme. More precisely, we
have to answer the question whether the joint probability
distribution $P(A,B)$, coming from the measurements performed by
Alice and Bob during the protocol, allow them to conclude that
the effectively distributed state was entangled or not. In
principle any separability criteria (see, e.g.,
\cite{Separability} and references therein) might be employed to
deliver this entanglement proof. The important question here is
whether the chosen criterion can can be used to provide a
necessary and sufficient condition to detect entanglement when
the knowledge about the state is not tomographic complete. As we
will see below, it is a property of EWs that they allow to obtain
a necessary and sufficient criterion for separability even when
the state cannot be completely reconstructed \cite{Curty04}.

Let us first consider EB schemes. In these schemes, Alice and Bob
perform some measurements on a bi-partite quantum state
distributed by an, in general, untrusted third party and retrieve
the probability distribution $P(A,B)$ of the outcomes. Before
showing that in this scenario EWs are specially appropriated to
detect entanglement, let us recall some facts about witnesses
\cite{Horodecki96,Terhal00,Lewenstein00}.

A witness is a Hermitean observable $W$ with a positive
expectation value on all separable states. So if a state $\rho$
obeys $Tr(\rho W)<0,$ the state $\rho$ must be entangled. We say
then that the state $\rho$ is detected by $W.$ In general, for
every entangled state there exists a witness detecting it,
however, this witness is in most cases very difficult to
construct. Witnesses can be {\it optimized} in the following
sense: A witness $W_1$ is called  {\it finer} than another
witness $W_2$ if $W_1$ detects all the states which are detected
by $W_2$ and some states in addition. Finally, a witness $W$ is
called {\it optimal,} when there is no other witness which is
finer than $W$ \cite{Lewenstein00}. Now we can state the
following (see also \cite{Curty04}):

\begin{theo}\label{necsuf}
Assume that Alice and Bob can perform some local measurements
with POVM elements $A_i\otimes B_i$, $i=1,...,n$, to obtain the
probability distribution of the outcomes $P(A,B)$ on the
distributed state $\rho.$ Then the correlations $P(A,B)$ cannot
originate from a separable state if and only if there is an EW of
the form $W=\sum_{i} c_{i}\ A_i\otimes B_i$ which detects the
effectively  distributed state, i.e., $Tr(W\rho)=\sum_{i} c_{i}\
P(A_i,B_i) < 0$.
\end{theo}

{\it Proof:} One direction of the above theorem is clear: If such
a witness with the properties from above exists, then the
effectively distributed state is clearly entangled. To prove the
other direction, let us look at the the following map, which maps
a quantum state $\rho$ to a real vector $\fA(\rho)\in{\mathbb
R}^n$,
\begin{equation}
\fA{}:\ \rho \mapsto \fA(\rho) = \{\fA(\rho)_0, \ldots,
\fA(\rho)_n\},
\end{equation}
where $\fA(\rho)_i=P(A_i,B_i)=Tr(A_i\otimes B_i\ \rho)$. That is,
it maps a state onto the set of probabilities or expectation
values of the POVM elements. This map is linear and, in general,
not injective. It maps the convex set $S$ of separable states onto
the convex set $S':=\fA(S).$  An entangled state $\rho$ with the
property $\fA(\rho)\in S'$ cannot be detected with the given
probabilities, since then there is a separable state $\rho_s$
being mapped to the same $\fA(\rho_s)=\fA(\rho),$ thus $\rho$ and
$\rho_s$ are indistinguishable. So a state $\rho_e$ for which
$P(A,B)$ cannot origin from a separable state must obey
$\fA(\varrho_e)\notin S',$ Now we have the usual construction of
witnesses: There must exist a hyper-plane separating
$\fA(\varrho_e)$ from $S'.$ This means that there is a vector
$w=(w_1,...,w_n)$ with $\sum_i w_i\ \fA(\varrho_e)_i < 0$ while
$\sum_i w_i\ \fA(\varrho)_i > 0$ for all $\rho$ with
$\fA(\rho)\in S'$. The observable $W=\sum_i w_i\ A_i\otimes B_i$
is now the desired EW, since $Tr(W\rho)=\sum_i w_i\ P(A_i,B_i).$
$\blacksquare$

We refer to witnesses that can be evaluated with the given POVM
elements and the corresponding correlations $P(A,B)$ as {\em
accessible}. According to Theorem \ref{necsuf}, the set of all
accessible witness operators gives  rise to a necessary and
sufficient condition for verifiable entanglement contained in the
correlations $P(A,B)$: The joint probability distribution
$P(A,B)$ can come exclusively from an entangled state if and only
if at least one accesible witness in the set gives rise to a
negative expectation value when it is evaluated with $P(A,B)$. Of
course, in this set there is some redundancy. Typically,  it
contains witnesses that are finer than others, and therefore one
can construct smaller sets of witnesses that are accessible and
still have the property of being necessary and sufficient for
verifying entanglement. Whenever this property holds, we refer
to  a set of witnesses   $\cal{W}$ as being a  {\em verification
set}. The ultimate goal will be to obtain a {\em minimal
verification set} in a compact description that contains no
further redundancies to allow an efficient systematic search for
verifiable entanglement by evaluating the members of this set.
The rest of this paper is mainly concerned with the search of
these minimal verification sets, although in the case of the
$4$-state P\&M protocol and in the $2$-state P\&M protocol we
find only {\em reduced verification sets}, which still may
contain some redundancies.

Before starting our quest for minimal verification  sets, let us
consider the case of P\&M schemes since in this section we have
considered, so far, only EB schemes. As we mentioned previously,
in this kind of schemes the reduced density matrix of Alice is
fixed since Eve has no access to the state of Alice to try to
modify it. However also this situation can be incorporated in the
theorem from above: We can add to the observables $A_i \otimes
B_i$ other observables $C_i\otimes \openone$ such that the
observables $C_i$ form a tomographic complete set of Alice's
Hilbert space. Those witnesses that can be evaluated with this
combined set of measurements can clearly be evaluated with the
measurements $A_i \otimes B_i$ and the knowledge of the reduced
density matrix of Alice.

In the geometric picture obtained in the proof of the theorem from
above, the knowledge of the expectation value of some of the
observables implies that we know that our state lies on some
hyperplane in the space of all expectation values. Then, we want
to decide for a point on this hyperplane whether is is in $S'$ or
not, and this can be done by witnesses. The knowledge of the mean
values of some observables may be used to argue that only a
smaller set of witnesses is relevant for such a P\&M scheme. We
will see an example of this later.

Finally, let us emphasize again that there are many other
separability criteria besides EWs which might be used for the
detection of entanglement in quantum cryptographic schemes. For
instance, the security of the first EB scheme proposed by Ekert in
$1991$ \cite{Ekert91}, the $4$-state EB scheme, was based on the
detection of quantum correlations by looking at possible
violations of Bell inequalities \cite{bellpaper}. This criterion,
or for example those based on uncertainty relations
\cite{hofmann1}, is directly linked to experimental data, which
makes the implementation simple. Another interesting criterion
that seems to be particularly suited for the case of P\&M
schemes, where the reduced density matrix of Alice is fixed and
known, is, for instance, the reduction criterion \cite{Mhoro99}.
However, it is not clear whether these criteria guarantee to
detect {\it all} entangled states which can be detected with the
given set of measurements. In fact, in the case of the $4$-state,
the knowledge of the performed measurements together with
$P(A,B)$ allows to detect entangled states beyond those that
violates Bell-like inequalities.

\section{QKD PROTOCOLS} \label{PROT}

We will now illustrate the consequences of this view for some
well-known QKD protocols. First we start reviewing the recent
results obtained in Ref.~\cite{Curty04} for the $6$-state and the
$4$-state EB protocols \cite{Bruss98,Ekert91}, which include a
minimal verification set to detect quantum correlations in both
protocols. Then we present the analysis for the case of P\&M
schemes. We investigate the $4$-state and the $2$-state P\&M
schemes \cite{BB84,Ben92}, and we obtain a reduced verification
set for each of these protocols. Finally, the last part of the
section gives a brief outlook to the study of quantum
correlations in higher dimensional QKD schemes and in practical
QKD \cite{jens04}.

\subsection{6-state protocol}

For the case of the 6-state EB protocol, Alice and Bob perform
projection measurements onto the eigenvectors of the three Pauli
operators $\sigma_x, \sigma_y,$ and $\sigma_z$ on the bi-partite
qubit states distributed by Eve. In the corresponding P\&M scheme
Alice prepares the eigenvectors of those operators by performing
the same measurements on a maximally entangled two-qubit state.
Note that here we are not using the general approach introduced
previoulsy, $|\Psi\rangle_{AB} =\sum_i \sqrt{p_i}
|\alpha_i\rangle|\varphi_i\rangle$, to model P\&M schemes, since
for this protocol it is sufficient to consider that the
effectively distributed quantum state consists only of two qubits.
In both cases Alice has complete tomographic knowledge of her
subsystem and therefore the class of EWs, which can be constructed
in both protocols, coincide. The set of three measurement bases
used in the protocol allows Alice and Bob to construct any EW of
the form
\begin{equation}\label{general}
W=\sum_{i,j=\{0,x,y,z\}} c_{ij}\ \sigma_i\otimes\sigma_j,
\end{equation}
where $\sigma_0=\openone$ and $c_{ij}$ are real numbers. Note that
the set of operators $\{\sigma_i\otimes\sigma_j\}_{i,j}$
constitutes an operator basis in
$\mathbb{C}^2\otimes{}\mathbb{C}^2$. This means that Alice and Bob
can in principle evaluate all EWs, in particular, the class of
optimal witnesses for two-qubit states. This class, denoted by
OEW, is given by the witnesses operators of the form
\cite{Lewenstein97}
\begin{equation}\label{optEW2}
W=|\phi_e\rangle\langle\phi_e|^{T_P},
\end{equation}
where $|\phi_e\rangle$ denotes any entangled state of two-qubit
systems and ${T_P}$ is the partial transposition, that is, the
transposition with respect to one of the subsystems
\cite{PartialTrans}. Therefore, in the $6$-state protocol, both
for EB and P\&M schemes, all entangled states can be detected and
the optimal witnesses OEW form the minimal verification set.

Alternatively to the witness approach, Alice and Bob can employ
as well quantum state tomography techniques to reconstruct the
effectively distributed quantum state and then use the
Peres-Horodecki criterion \cite{Peres96,Horodecki96} to determine
whether that state was entangled or not. This criterion
establishes that a two-qubit state is separable iff its partial
transposition is positive. For higher dimensional systems,
however, note that although all operators with non-positive
partial transposition are entangled, there exist PPT entangled
states \cite{Phoro97}.

\subsection{4-state protocol}
While the analysis of the $6$-state protocol is quite simple, due
to the complete tomographic information that Alice and Bob share,
the $4$-state protocol needs a deeper examination. As we will
show below, the class of OEW for two-qubit systems cannot be
evaluated with the given correlations neither in the EB nor in the
P\&M versions of the protocol. In the EB case Alice and Bob
perform projection measurements onto two mutually unbiased bases,
say the ones given by the eigenvectors of the two Pauli operators
$\sigma_x$ and $\sigma_z$. In the corresponding P\&M scheme,
Alice can use as well the same set of measurements but now on a
maximally entangled state. Here again, like in the $6$-state
protocol, we use the fact that the approach $|\Psi\rangle_{AB}
=\sum_i \sqrt{p_i} |\alpha_i\rangle|\varphi_i\rangle$ to model
P\&M schemes can be reduce to employ only two-qubit states for
this protocol. Let's begin our analysis for the EB scheme
\cite{Curty04}.

\subsubsection{Entanglement-based (EB)}

In the case of the $4$-state EB protocol we will denote the set
of EWs that can be evaluated with the resulting correlations as
W$^{EB}_4$. All elements are of the form
\begin{equation}\label{4QKD}
W^{EB}_4=\sum_{i,j=\{0,x,z\}} c_{ij}\ \sigma_i\otimes\sigma_j.
\end{equation}
This class of EWs can be characterized with the following
observation.
\begin{lem}
\cite{Curty04} Given an entanglement witness $W$ we find
$W\in{}W^{EB}_4$ iff $W=W^{T}=W^{T_P}$.
\end{lem}
{\it Proof:} To see this, we start with the general Ansatz of
Eq.~(\ref{general}) and we impose the conditions
$W=W^{T}=W^{T_P}$. This directly constraints $W$ to the form
(\ref{4QKD}) since $\sigma_y$ is the only skew-symmetric element
in the operator basis. The reverse direction is then trivial.
$\blacksquare$

It is straightforward to see that the elements of OEW do not
fulfill this condition. Below, we will provide a criterion to
decide if an entangled state can be detected by $W\in{}W^{EB}_4$.
This means that, in contrast to the case of the $6$-state
protocol, in the $4$-state EB protocol there can be entangled
states that give rise to correlations $P(A,B)$ that are not
sufficient to prove the presence of entanglement.

The concept of optimal witnesses introduced in Section \ref{EW}
for general witness operators can as well be extended to the
witnesses which are accessible with the given set of
measurements. This way we call a witness $W$ {\em optimal in
class $C$} iff there is no other element in $C$ that detects all
entangled states detected by $W$. Our goal now is to characterize
a complete family of witness operators that are optimal in the
class W$^{EB}_4$. This family forms the minimal verification set.
Then it is sufficient to check this family to decide whether the
presence of entanglement can be verified from the given data. To
do this we start presenting a necessary and sufficient condition
for a bi-partite state to contain entanglement that can be
detected by elements of W$^{EB}_4$.
\begin{lem}
\label{symmetry} \cite{Curty04} An entangled state $\rho$ can be
detected with a $W\in{}W^{EB}_4$ iff the operator
$\Omega=\frac{1}{4}\left(\rho+\rho^{T_A}+\rho^{T_B}+\rho^T\right)$
is a non-positive operator.
\end{lem}
{\it Proof:} To see this, let us start by the observation that the
symmetries of the witness operators in W$^{EB}_4$ give rise to
the identity ${\rm Tr}\left(W \rho \right) = {\rm Tr}\left(W
\Omega\right)$. Now let us assume that the operator $\Omega$ is
non-negative. Then one can interpret it as a density matrix.
Since it is invariant under partial transposition, it must be a
separable state. Since $W$ is a witness operator, we must
therefore find ${\rm Tr}\left(W \rho \right) \geq 0$. As a
result, we find that the non-positivity of $\Omega$ is a
necessary condition to detect entanglement of the state $\rho$
with witnesses in W$^{EB}_4$. The reverse direction is included
here only for completeness and the proof is included implicitly
in Theorem \ref{theo4EB}. $\blacksquare$

Next we present a set of EWs composed of optimal witnesses in the
class W$^{EB}_4$ which forms a minimal verification set of the
$4$-state EB protocol.

\begin{theo}\label{theo4EB}
\cite{Curty04} Consider the family of operators
$W=\frac{1}{2}(Q+Q^{T_P})$, where
$Q=|\phi_e\rangle\langle\phi_e|$ and $|\phi_e\rangle$ denotes a
real entangled state. The elements of this family are witness
operators that are optimal in W$^{EB}_4$ (OEW$^{EB}_4$) and
detect all the entangled states that can be detected within
W$^{EB}_4$.
\end{theo}
{\it Proof:} Let us start by checking that this family, indeed,
can detect all entanglement that can be detected in W$^{EB}_4$.
{From} the Observation \ref{symmetry} we know that we need only to
consider bi-partite states $\rho$ such that $\Omega =
\frac{1}{4}\left(\rho+\rho^{T_A}+\rho^{T_B}+\rho^T\right)$ is
non-positive. We have, therefore, that there exists always an
(entangled) state $|\phi_e\rangle$ such that
$\langle\phi_e|\Omega|\phi_e\rangle<0$. Moreover, since
$\Omega=\Omega^T$, this operator has a real representation. In
this representation, also the state $|\phi_e\rangle$ has a real
representation \cite{Strang80}. Let us define the projector
$Q=|\phi_e\rangle\langle\phi_e|$. Then we find
$\langle\phi_e|\Omega|\phi_e\rangle= {\rm
Tr}\left(\frac{1}{4}\left(Q+Q^{T_A}+Q^{T_B}+Q^T\right)\rho
\right)$. This means that we can define the operator $W=
\frac{1}{4}\left(Q+Q^{T_A}+Q^{T_B}+Q^T\right)$ that can be further
simplified to $W=\frac{1}{2}\left(Q+Q^{T_P}\right)$ thanks to the
real representation of $Q$. This operator is a witness operator,
since ${\rm Tr}\left(W \sigma \right)\geq 0$ for all separable
states $\sigma$, while ${\rm Tr}\left(W \rho \right)< 0$ for the
chosen $\rho$. Moreover, by construction the family of these
witness operators detects all entanglement that can be detected
within W$^{EB}_4$.

Finally, we need to show that all witnesses of this new set
$W=\frac{1}{2}\left(Q+Q^{T_P}\right)$ are optimal within
W$^{EB}_4$ so they form OEW$^{EB}_4$. In Ref.~\cite{Lewenstein00}
it has been proven that, given a set of witness operators S$_W$,
$W\in{}$S$_W$ is optimal in S$_W$ iff for all positive
semi-definite operators $P$ and $\epsilon>0$,
$W'=(1+\epsilon)W-\epsilon{}P{}\notin{}$S$_W$. When a $P$ can be
subtracted, it has to fulfill $\langle{}e,f|P|e,f\rangle={0}$ for
all product vectors
$|e,f\rangle{}\in\mathbb{C}^2\otimes{}\mathbb{C}^2$ with
$\langle{}e,f|W|e,f\rangle=0$ since otherwise we would not have a
witness anymore. In the case of witness operators of the form
$W=\frac{1}{2}(Q+Q^{T_P})$, where
$Q=|\phi_e\rangle\langle\phi_e|$ and
\begin{equation}
|\phi_e\rangle=\sum_{i=0}^1{}c_i|i\rangle|i\rangle
\end{equation}
denotes the Schmidt decomposition of $|\phi_e\rangle$, we have
that the $|e,f\rangle$ that satisfy $\langle{}e,f|W|e,f\rangle=0$
are given by $|0\rangle|1\rangle$, $|1\rangle|0\rangle$, and the
unnormalized states
\begin{equation}
|\phi{}(\lambda)\rangle=(\lambda|0\rangle\pm\sqrt{1-\lambda^2}|1\rangle)
(c_1\sqrt{1-\lambda^2}|0\rangle\mp{}c_0\lambda|1\rangle),
\end{equation}
with $\lambda\in(0,1)$. These product vectors span a three
dimensional subspace that is orthogonal to $|\phi_e\rangle$. This
means that $P$ cannot be subtracted from $W$ unless $P=Q$. But
$(1+\epsilon)W-\epsilon{}Q{}=
\frac{1}{2}[(1-\epsilon)Q+(1+\epsilon)Q^{T_P}]\notin{}$W$^{EB}_4$
for all $\epsilon>0$. Therefore all witness operators
$W=\frac{1}{2}(Q+Q^{T_P})$ with $Q=|\phi_e\rangle\langle\phi_e|$
and $|\phi_e\rangle$ real are OEW$^{EB}_4$'s. $\blacksquare$

\subsubsection{Prepare\&Measure (P\&M)}

Once we have presented a set of witness operators that is optimal
for the EB scheme, we will show below that this family is also
sufficient to detect all entangled states that can be detected in
the P\&M version of the $4$-state protocol. That is, with respect
to the ability to detect quantum correlations, both schemes can
use the same verification set. As we showed in Section \ref{EW},
in the case of P\&M schemes one can add to the set of observables
measured in the protocol other observables $C_i\otimes \openone$
such that the observables $C_i$ form a tomographic complete set
of Alice's Hilbert space. So we have to add the operator
$\sigma_y\otimes\sigma_0$ to the observables in Eq.~(\ref{4QKD}).
This way one obtains all the witnesses that can be evaluated in
the $4$-state P\&M protocol. This new set, that we shall denote
as W$^{P\&M}_4$, is of the form
\begin{equation}\label{4QKD_PM}
W^{P\&M}_4=\sum_{i,j=\{0,x,z\}} c_{ij}\
\sigma_i\otimes\sigma_j+c_{y0}\ \sigma_y\otimes\sigma_0 \; .
\end{equation}
where $c_{y0}$ is as well a real number.

Now we present a result, Observation \ref{ObsEW}, that applies to
the observables given by Eq.~(\ref{4QKD}) and that will be useful
to prove that the set of OEW$^{EB}_4$ obtained in Theorem
\ref{theo4EB} is also sufficient to detect all entangled states
that can be detected in the $4$-state P\&M scheme and, therefore,
it forms a reduced verification set of this protocol.

\begin{lem}\label{ObsEW}
Given an observable $W$ with $W=W^{T}=W^{T_P}$, then ${\rm
Tr}(W{}\sigma)\geq{}0$ for all $\sigma$ separable iff ${\rm
Tr}(W{}\sigma_r)\geq{}0$ for all $ \sigma{}_r$ real and separable.
\end{lem}
{\it Proof:} (If) Using the fact that $W=W^{T}$ we have ${\rm
Tr}(W\sigma)=\frac{1}{2}{\rm Tr}[W(\sigma+\sigma^T)]$. Note that
the symmetric matrix $\sigma+\sigma^T$ is real and positive
semi-definite because $\sigma$ is hermitian and positive
semi-definite and transposition is a positive operation. Moreover
$(\sigma+\sigma^T)^{T_P}$ is positive semi-definite since
$\sigma^{T_P}$ is positive semi-definite for all $\sigma$
separable. This means that $\frac{1}{2}(\sigma+\sigma^T)$ is a
real separable quantum state. Therefore if ${\rm
Tr}(W{}\sigma_r)\geq{}0$ for all $ \sigma{}_r$ real and separable
then ${\rm Tr}(W\sigma)\geq{}0$ for all $\sigma$ separable. (Only
if) The proof is trivial. $\blacksquare$

\begin{theo}
The family of OEW$^{EB}_4$, $W=\frac{1}{2}(Q+Q^{T_P})$ with
$Q=|\phi_e\rangle\langle\phi_e|$ and $|\phi_e\rangle$ a real
entangled state, is sufficient to detect all entangled states
that can be detected in the $4$-state P\&M scheme.
\end{theo}

{\it Proof:} To be EWs, the operators W$^{P\&M}_4$ given by
Eq.~(\ref{4QKD_PM}) must satisfy ${\rm
Tr}(W^{P\&M}_4{}\sigma)\geq{}0$ for all $\sigma$ separable. In
particular, it must satisfy ${\rm
Tr}(W^{P\&M}_4{}\sigma_r)\geq{}0$ for all $ \sigma{}_r$ real and
separable. We have that the term $\sigma_y\otimes\sigma_0$
satisfies
\begin{equation}
Tr[(\sigma_y\otimes\sigma_0) \sigma_r]=0,\ \forall\ \sigma{}_r.
\end{equation}
This means, therefore, that we need to guarantee that the first
term in Eq.~(\ref{4QKD_PM}) satisfies
\begin{equation}
\sum_{i,j=\{0,x,z\}} c_{ij}\ Tr[(\sigma_i\otimes\sigma_j)
\sigma_r]\geq{}0,\ \forall\ \sigma{}_r.
\end{equation}
According to Observation \ref{ObsEW}, we obtain that the term
$\sum_{i,j=\{0,x,z\}} c_{ij}\ \sigma_i\otimes\sigma_j$ in
Eq.~(\ref{4QKD_PM}) has to be an EW which belongs to the class
W$^{EB}_4$. That is, W$_4^{P\&M}=$W$^{EB}_4+c_{y0}\
\sigma_{y}\otimes\sigma_0$.

To conclude the proof, now we have to take into account that in
the $4$-state P\&M scheme the reduced density matrix of Alice is
fixed and given by $\rho_A=\frac{1}{2}\openone{}$. This means that
\begin{equation}
Tr(W^{P\&M}_4 \rho)=Tr(W^{EB}_4 \rho)
\end{equation}
for all $\rho$ such as $Tr_B(\rho)=\frac{1}{2}\openone{}$, since
$Tr[(\sigma_y\otimes\sigma_0) \rho]=0$. That is, the entangled
states that can be detected in the P\&M protocol are also detected
by the class of witnesses W$^{EB}_4$. We have, therefore, that is
sufficient to consider the set of OEW$^{EB}_4$. $\blacksquare$

\subsubsection{Evaluation}

From the set of witness operators OEW$^{EB}_4$ given by
$W=\frac{1}{2}(Q+Q^{T_P})$, with $Q=|\phi_e\rangle\langle\phi_e|$
and $|\phi_e\rangle$ a real entangled state, one can obtain a
necessary and sufficient condition for the presence of
entanglement in the observable correlations $P(A,B)$. This result
applies to both versions of the $4$-state protocol: EB and P\&M.
Note that for the case of an EB scheme all the witnesses in
OEW$^{EB}_4$ are optimal and form a minimal verification set.
However, for a P\&M scheme some of them might be redundant and
therefore this set of EWs forms a reduced verification set for
this version of the protocol. The set OEW$^{EB}_4$ includes an
infinity number of witness operators, but, as we will see below,
these EWs can be easily parametrized with only three real
parameters. From a practical point of view, this means that Alice
and Bob can easily check the conditions $Tr(W\ \rho)$ with $W\in$
OEW$^{EB}_4$ numerically.

Let us briefly analyze the implications of our results in the
relationship between the bit error rate $e$ in the $4$-state and
in the $6$-state protocols and the presence of correlations of
quantum mechanical nature \cite{Curty04}. Here the error rate $e$
quantifies the rate of events where Alice and Bob obtain
different results. It refers to the sifted key, i.e considering
only those cases where the signal preparation and detection
methods employ the same polarization basis. In an intercept-resend
attack Eve measures every signal emitted by Alice and prepares a
new one, depending on the result obtained, that is given to Bob.
This action corresponds to an entanglement breaking channel
\cite{Horodecki03}, i.e., it is a channel $\Phi$ such as
$I\otimes{}\Phi(\rho)$ is separable for any density matrix $\rho$
on a tensor product space. Such a channel gives rise to $e\geq
25\%$ ($4$-state protocol) and $e \geq 33\%$ ($6$-state
protocol), respectively \cite{Ekert94,Bruss98}, which might seem
to indicate that these values represent an upper bound for the
tolerable error rate in the protocols (see also \cite{niko}).
However, it turns out that for some asymmetric error patterns, it
is possible to detect the presence of quantum correlations even
for error rates above $25\%$ ($33\%$) \cite{Curty04}. Let us
illustrate this fact with two examples that are motivated by the
propagation of the polarization state of a single photon in an
optical fiber.

In the first example we will consider a channel described by an
unitary transformation that changes on a timescale much longer
than the repetition cycle of the signal source, so it can be
thought to be constant over that time: for instance, the channel
given by the unitary transformation
$U(\theta)=\cos\theta\openone-i\sin\theta\sigma_y$. In this
scenario, the resulting distributed state for both QKD protocols
is given by $|\psi\rangle=\cos\theta|00\rangle+
\sin\theta|01\rangle-\sin\theta|10\rangle+\cos\theta|11\rangle$.
The corresponding bit error rate is $e=\sin^2\theta$ and
$e=\frac{2}{3}\sin^2\theta$ for the $4$-state and the $6$-state
protocols, respectively. Nevertheless, it can be shown that in
both cases the existence of quantum correlations can be detected
for all angles $\theta$. The case of the $6$-state protocol is
clear, since a unitary transformation preserves the entanglement
and all entanglement can be verified in this protocol. With
respect to the $4$-state protocol, note there is always an
entanglement witness $W\in$ OEW$^{EB}_4$ that detects quantum
correlations in $P(A,B)$. In particular, let us use
$W=\frac{1}{2}(|\phi_e\rangle\langle\phi_e|+
|\phi_e\rangle\langle\phi_e|^{T_P})$, with $|\phi_e\rangle$ as the
eigenvector of the operator
$\frac{1}{2}|\psi\rangle\langle\psi|^{T_P}$ which corresponds to
its negative eigenvalue. Then we find in a suitable representation
as a {\it pseudo-mixture} \cite{Sanpera97} for the entanglement
witness that ${\rm Tr} (W \rho) = \sum_i{}c_i{}\
P(a_i,b_i)=-\frac{1}{4}$.

\begin{figure}
\begin{center}
\includegraphics[scale=.4]{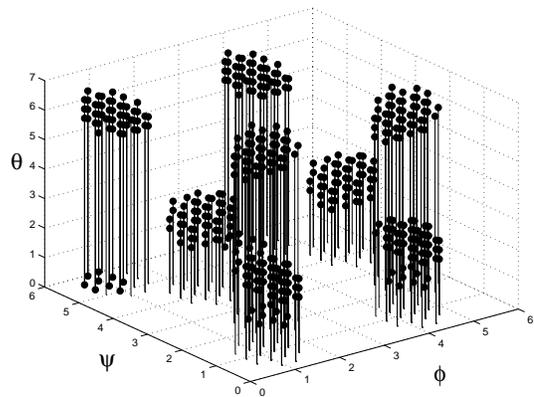}
\end{center}
\caption{Illustration of several regimes of the parameters $\phi,
\psi$, and $\theta$ leading to negative expectation values of the
operators
OEW$^{EB}_4=\frac{1}{2}(|\phi_e\rangle\langle\phi_e|+|\phi_e\rangle\langle\phi_e|^{T_P})$,
with
$|\phi_e\rangle=\cos\phi|00\rangle+\sin\phi(\cos\psi|01\rangle+\sin\psi(\cos\theta|10\rangle+\sin\theta|11\rangle))$,
when they are evaluated on the joint probability distribution
$P(A,B)$ given in Table~\ref{table4}.\label{Fig4}}
\end{figure}

For the second example, we focus on the $4$-state EB protocol
only and we consider the particular joint probability distribution
$P(A,B)$ given by Table~\ref{table4}, where the states
$|\pm\rangle$ are defined as
$|\pm\rangle=1/\sqrt{2}(|0\rangle\pm{}|1\rangle)$. In principle,
it is not straightforward to decide whether these correlations
can be explained as coming exclusively from an entangled state or
not. This is specially so since in this case the resulting bit
error rate is given by $e\approx{}35.4\%$.
\begin{table}
\begin{tabular}{|c|c|c|c|c|}
\hline\hline
    $A\setminus{}B$ & $|0\rangle$ & $|1\rangle$ & $|+\rangle$ & $|-\rangle$\\
\hline $|0\rangle$ & 0.08058 & 0.04757 & 0.02106 & 0.10709 \\
$|1\rangle$ & 0.04623 & 0.07560 & 0.11349 & 0.00834 \\
$|+\rangle$ & 0.11808 & 0.01690 & 0.09319 & 0.04179 \\
$|-\rangle$ & 0.00873 & 0.10627 & 0.04136 & 0.07364\\
\hline\hline
\end{tabular}
\caption{Example of a $P(A,B)$ for the $4$-state EB protocol,
where $|\pm\rangle=1/\sqrt{2}(|0\rangle\pm{}|1\rangle)$. The
table is normalized such as $\sum_i P(A_i,B_i)=1$. \label{table4}}
\end{table}
To decide that question systematically we can use the
verification set defined by OEW$^{EB}_4$:
$W=\frac{1}{2}(|\phi_e\rangle\langle\phi_e|+|\phi_e\rangle\langle\phi_e|^{T_P})$.
The real states $|\phi_e\rangle$ can be parametrized as
$|\phi_e\rangle=\cos\phi|00\rangle+\sin\phi(\cos\psi|01\rangle+\sin\psi(\cos\theta|10\rangle+\sin\theta|11\rangle))$,
with only three real parameters $\phi, \psi,
\theta\in{}[0,2\pi)$. Moreover, since the state $|\phi_e\rangle$
is entangled these parameters satisfy additionally the condition:
$\sin\phi\sin\psi(\sin\phi\cos\psi\cos\theta-\cos\phi\sin\theta)\neq{}0$
\cite{Wootters98}. However, from a practical point of view it
might be easier just to consider all angles $\phi, \psi$, and
$\theta$ and allow the evaluation of some positive operators.
After expressing the witness operators as a {\it pseudo-mixture}
\cite{Sanpera97} the condition $Tr(W \rho)<0$ can be rewritten as
$\sum_i f_i(\phi,\psi,\theta)\ P(A_i,B_i)<0$, with
$c_i=f_i(\phi,\psi,\theta)$ for some functions $f_i$. Now it is
easy to search numerically through the space of parameters $\phi,
\psi$, and $\theta$ for quantum correlations in $P(A,B)$. This
fact is illustrated in Figure~\ref{Fig4}, where some combinations
of these parameters detecting entanglement when they are
evaluated on the $P(A,B)$ given in Table~\ref{table4} are marked.

To finish this section we show now that the family of witness
operators OEW$^{EB}_4$ allow to detect entangled states beyond
those that violates Bell-like inequalities \cite{bellpaper}. Note
that, as we mentioned previously, the security of the original
$4$-state EB scheme \cite{Ekert91} was based on the detection of
entanglement by looking at possible violations of Bell
inequalities, which is in principle more restrictive than
detection of quantum correlations. It is known that a violation
of a Bell inequality can be formally expressed as an EW
\cite{Terhal00}, while the contrary does not hold always. An
interesting question then is to ask whether the family of
OEW$^{EB}_4$ correspond to Bell inequalities or not. It is easy
to see that the knowledge of the performed measurements in the
$4$-state protocol together with the joint probability
distribution $P(A,B)$ allow to detect entangled states that do
not violate Bell-like inequalities. Consider for instance the
two-qubit entangled states introduced by Werner in
\cite{Werner89} and which are defined as
\begin{equation}
\rho_{W}=p|\psi^{-}\rangle\langle\psi^{-}|+\frac{1}{4}(1-p)\openone,
\end{equation}
with $|\psi^{-}\rangle=1/\sqrt{2}(|01\rangle-|10\rangle)$. In the
probability range $1/3<p<1/\sqrt{2}$ Werner states do not violate
any known Bell inequalities \cite{horo95} but nevertheless they
can be detected with the witnesses OEW$^{EB}_4$ for
$p>\frac{1}{2}$. To see this, note that the operator
$\Omega=1/4(\rho_{W}+\rho_{W}^{T_A}+\rho_{W}^{T_B}+\rho_{W}^T)$ is
a non-positive operator for $p>\frac{1}{2}$.

\subsection{2-state protocol}

The $2$-state protocol \cite{Ben92} is one of the simplest QKD
protocols, since it is based on the random transmission of only
two nonorthogonal states, $|\varphi_0\rangle$ and
$|\varphi_1\rangle$. Alice chooses randomly a bit value $i$, and
prepares a qubit in the state
$|\varphi_i\rangle=\alpha|0\rangle+(-1)^i\beta|1\rangle$, with
$0<\alpha<1/\sqrt{2}$ and $\beta=\sqrt{1-\alpha^2}$, that is sent
it to Bob. On the receiving side, Bob measures the qubit he
receives in a basis chosen, independently and at random, within
the set $\{\{|\varphi_0\rangle,|\varphi_0^{\perp}\rangle\},\{
|\varphi_1\rangle,|\varphi_1^{\perp}\rangle\}\}$, with
$|\langle\varphi_i|\varphi_i^{\perp}\rangle|=0$. Note that
alternatively to this detection method, Bob could also perform a
POVM defined by the operators
$F_0=|\varphi_1^{\perp}\rangle\langle\varphi_1^{\perp}|/2$,
$F_1=|\varphi_0^{\perp}\rangle\langle\varphi_0^{\perp}|/2$, and
$F_{null}=\openone-F_0-F_1$. Using the ideas introduced by
Bennett et al. \cite{Mermin92}, one can also think of the
preparation process in the following way: Alice prepares an
entangled bi-partite state of the form
\begin{equation}\label{B92}
|\Psi\rangle_{AB}=\frac{1}{\sqrt{2}}(|0\rangle_A|\varphi_0\rangle_B+|1\rangle_A|\varphi_1\rangle_B),
\end{equation}
and then she measures her subsystem in the basis
$\{|0\rangle,|1\rangle\}$. Note that in this scheme the fact that
the reduced density matrix of Alice is fixed and equal to
$\rho_A=Tr_B(|\Psi\rangle\langle\Psi|_{AB})$ with
$|\Psi\rangle_{AB}$ given by Eq.~(\ref{B92}) is vital to detect
quantum correlations in $P(A,B)$. Otherwise, the joint probability
distribution $P(A,B)$ alone does not allow Alice and Bob to
distinguish between the entangled state $|\Psi\rangle_{AB}$ and
the separable one $\sigma_{AB}=\frac{1}{2}\sum_{i=0}^1
|i\rangle\langle{}i|_A\otimes|\varphi_i\rangle\langle\varphi_i|$.

We obtain, therefore, that the set of EWs that can be evaluated
in the $2$-state protocol, and which we shall denote as W$_2$, is
of the form
\begin{equation}\label{2QKD}
W_2=\sum_{\begin{array}{c} _{i=\{0,z\}} \\_{j=\{x,z\}}
\end{array}} c_{ij}\ \sigma_i\otimes\sigma_j+\sum_{k=\{0,x,z,y\}}
c_{k}\ \sigma_k\otimes\sigma_0,
\end{equation}
where the second term in Eq.~(\ref{2QKD}) includes a set of
observables such as Alice can reconstruct completely the state of
her subsystem. This family of witness operators can equivalently
be rewritten as
\begin{equation}\label{Alt2}
W_2=|0\rangle\langle{}0|\otimes{}A+|1\rangle\langle{}1|\otimes{}B+x\
C(\theta),
\end{equation}
where $A$ and $B$ represent two real symmetric operators, $A=A^T$
and $B=B^T$, given by
\begin{equation}
A=\sum_{i=\{0,x,z\}} (c_{0i}+c_{zi})\ \sigma_{i}
\end{equation}
and
\begin{equation}
B=\sum_{i=\{0,x,z\}} (c_{0i}-c_{zi})\ \sigma_{i},
\end{equation}
respectively, the parameter $x$ is given by $x=|c_x+ic_y|\geq{}0$,
and
\begin{equation}\label{Ctheta}
C(\theta)=\left(\begin{array}{c|c}
0 & e^{i\theta}\ \openone \\
\hline e^{-i\theta}\ \openone & 0
\end{array}\right),
\end{equation}
with $\theta=\tan^{-1}(c_y/c_x)$. That is, we have included the
two observables $\sigma_x\otimes\openone$ and
$\sigma_y\otimes\openone$ that appear in Eq.~(\ref{2QKD}) in the
term $x\ C(\theta)$ of Eq.~(\ref{Alt2}), while the remaining
observables in Eq.~(\ref{2QKD}) are included in the first two
terms,
$|0\rangle\langle{}0|\otimes{}A+|1\rangle\langle{}1|\otimes{}B$,
of Eq.~(\ref{Alt2}).

It is straightforward to see that, as in the case of the $4$-state
(EB and P\&M) protocol, the class of witness operators $W_2$ does
not allow to evaluate the OEW given in Eq.~(\ref{optEW2}). Note
that the elements $W$ in OEW satisfy $W\ngeq{}0$ and
$W^{T_P}\geq{}0$. The witnesses in the class $W_2$, on the
contrary, fulfill $W_2=W_2^{T_B}$, which means that $W_2^{T_B}$
cannot be a positive semi-definite operator unless $W_2$ is also
positive semi-definite. Therefore, in the $2$-state protocol
there can be also entangled states that give rise to correlations
$P(A,B)$ that are not sufficient to prove the presence of
entanglement. In the same way, it is interesting to note also
that there are no witnesses in the set of $W_2$ that belongs to
the family of
OEW$^{EB}_4=\frac{1}{2}(|\phi_e\rangle\langle\phi_e|+|\phi_e\rangle\langle\phi_e|^{T_P})$.
To see this, note that the representation given in
Eq.~(\ref{Alt2}) is incompatible with the fact that the state
$|\phi_e\rangle$ is a real entangled state. In the rest of this
section we obtain a reduced verification set of the $2$-state
protocol (Theorem \ref{wit2}).

The first requirement that an operator of the form given by
Eq.~(\ref{Alt2}) must satisfy to be an EW is $Tr(W_2
\sigma)\geq{}0$ for all separable states $\sigma$. Since the set
of separable states is convex, this condition is equivalent to ask
$Tr(W_2
|\phi\rangle\langle\phi|_A\otimes|\psi\rangle\langle\psi|_B)\geq{}0$
for all states $|\phi\rangle_A|\psi\rangle_B$. Let us start by
considering states of the form $|0\rangle_A|\psi\rangle_B$. We
have that $Tr(W_2
|0\rangle\langle{}0|_A\otimes|\psi\rangle\langle\psi|_B)=\langle\psi|A|\psi\rangle$,
which means that the operator $A$ must be positive semi-definite,
i.e., $A\geq{}0$. In the same way, but now using the separable
pure states $|1\rangle_A|\psi\rangle_B$, one obtains $B\geq{}0$.
This means that the first two terms of Eq.~(\ref{Alt2}),
$|0\rangle\langle{}0|\otimes{}A+|1\rangle\langle{}1|\otimes{}B$,
represent a positive semi-definite operator and the only term
responsable to detect quantum correlations is the one given by
$x\ C(\theta)$. Note that this implies that the class $W_2$ does
not allow to detect maximally entangled states
$|\phi_e\rangle=1/\sqrt{2}\sum_{i=0}^1
|\psi_i\rangle|\varphi_i\rangle$, with
$\langle\psi_i|\psi_j\rangle=\langle\varphi_i|\varphi_j\rangle=\delta_{ij}$,
since in that case it turns out that
$\langle\phi_e|C(\theta)|\phi_e\rangle=0$. This fact is not too
surprising since the distributed states in this protocol are not
maximally entangled. Let us now come back to the general case:
$Tr(W_2
|\phi\rangle\langle\phi|_A\otimes|\psi\rangle\langle\psi|_B)\geq{}0$.
If we express the state $|\phi\rangle_A$ as
$|\phi\rangle_A=\alpha|0\rangle+\beta|1\rangle$, this condition
reduces to
\begin{equation}\label{x1}
2|\alpha\beta|x\leq{}\langle\psi|(|\alpha|^2A+|\beta|^2B)|\psi\rangle,\
\ \forall\ |\psi\rangle
\end{equation}
and for all $\alpha$ and $\beta$ such that
$|\alpha|^2+|\beta|^2=1$. After optimizing over the parameters
$\alpha$ and $\beta$, Eq.~(\ref{x1}) can be further simplified to
\begin{equation}\label{minx}
x\leq{}\textrm{min}_{|\psi\rangle}\sqrt{\langle\psi|A|\psi\rangle\langle\psi|B|\psi\rangle}.
\end{equation}
That is, whenever the value of the parameter $x$ is below the
bound given by Eq.~(\ref{minx}), $W_2$ has a positive expectation
value on all separable states.

Now we will provide a necessary and sufficient condition for the
operators $W_2$ in order to detect entanglement, i.e, to
guarantee $W_2\ngeq{}0$. First, it is straightforward to see that
the operators $A$ and $B$ have to be of full rank, otherwise
according to Eq.~(\ref{minx}) we have $x=0$ and
$W_2=|0\rangle\langle{}0|\otimes{}A+|1\rangle\langle{}1|\otimes{}B\geq{}0$.
Now we can prove the following observation:
\begin{lem}\label{positive}
An operator
$W_2=|0\rangle\langle{}0|\otimes{}A+|1\rangle\langle{}1|\otimes{}B+x\
C(\theta)$, with $A$ and $B$ being two real symmetric positive
operators, $A>0$ and $B>0$, and the operator $C(\theta)$ given by
Eq.~(\ref{Ctheta}) with $\theta\in{}[0,2\pi)$ satisfies
$W_2\ngeq{}0$ iff $x>x_{min}$ with
\begin{equation}
x_{min}=\sqrt{\frac{\alpha}{2}-\sqrt{\frac{\alpha^2}{4}-{det(A)}{det(B)}}},
\end{equation}
and where $\alpha=Tr(AB)$.
\end{lem}
{\it Proof:} See Appendix \ref{A}.

\begin{theo}\label{wit2}
The family of witness operators
$W_2=|0\rangle\langle{}0|\otimes{}A+|1\rangle\langle{}1|\otimes{}B+x\
C(\theta)$, with $A$ and $B$ being two real symmetric positive
operators, $A>0$ and $B>0$, the operator $C(\theta)$ given by
Eq.~(\ref{Ctheta}) with $\theta\in{}[0,2\pi)$, and such as
$x=\textrm{min}_{|\psi\rangle}\sqrt{\langle\psi|A|\psi\rangle\langle\psi|B|\psi\rangle}>x_{min}$
is sufficient to detect all entangled states that can be detected
in the $2$-state protocol.
\end{theo}
{\it Proof:} According to the results presented above, we only
need to prove that given a witness operator $W_2\ngeq{}0$ with a
value of $x$ that saturates the bound of Eq.~(\ref{minx}), and
that we shall denote as $W_2(x_{max})$, is finer than the same
witness $W_2(x)$ with an $x<x_{max}$. That is,
\begin{equation}\label{opt}
Tr[W_2(x_{max})\rho]\leq{}Tr[W_2(x)\rho]
\end{equation}
for all $\rho$ entangled and detected by $W_2(x)$. Since
$W_2(x_{max})$ and $W_2(x)$ share the same operators $A$ and $B$
by definition, Eq.~(\ref{opt}) can be further simplify to
\begin{equation}
(x_{max}-x)\ Tr[C(\theta)\rho]\leq{}0.
\end{equation}
Finally note that this condition is always fulfilled, since
$(x_{max}-x)>0$ and $Tr[C(\theta)\rho]<0$, otherwise $\rho$
cannot be detected by $W_2(x)$. $\blacksquare$

The coefficients of the pseudo-mixture decomposition of the
witness operators given by Theorem \ref{wit2} can be parametrized
in this case with six real parameters.

\subsection{Higher dimensional QKD protocols}

So far we have searched for quantum correlations in qubit-based
QKD protocols which means we have restricted ourselves to
operators in $\mathbb{C}^2\otimes{}\mathbb{C}^2$. This fact makes
the characterization of a given class of witness operators
easier, since for systems defined in
$H=\mathbb{C}^2\otimes{}\mathbb{C}^2$ and
$H=\mathbb{C}^2\otimes{}\mathbb{C}^3$ all witness operators
belong to the class of so-called {\it decomposable entanglement
witnesses} (DEWs) which has a simple well-known form:
$W=\epsilon{}P+(1-\epsilon)Q^{T_P}$, with $P\geq{}0$ and
$Q\geq{}0$ satisfying $Tr(P)=Tr(Q)=1$, and $\epsilon\in[0,1)$
\cite{woro76}. In the case of systems of higher dimension than
those in $H=\mathbb{C}^2\otimes{}\mathbb{C}^2$ and
$H=\mathbb{C}^2\otimes{}\mathbb{C}^3$ not all the witness
operators are DEWs; for example DEWs cannot detect PPT entangled
states \cite{Phoro97}. It is necessary to use also the so-called
{\it non-decomposable entanglement witnesses} (NDEWs). This fact
makes the characterization of witness operators more subtle and
so far it is still not clear how to construct such witness
operators even when the information about the state is
tomographic complete \cite{ndew,Terhal00,Lewenstein00}. As we
mentioned already before, ideally the goal is to obtain a compact
description of the minimal verification set for a given QKD
protocol in order to systematically search for entanglement in
$P(A,B)$. However, due to the fact that the characterization of
NDEWs is not easy to handle it might be of interest to obtain, at
least, ${\it one}$ relevant EW within the proper class as a first
step towards the demonstration of successful QKD.

Recently, it has been shown that a good deal of new insight in
the optimization of NDEWs can come from the theory of convex
optimization \cite{Doherty,Brasilians,jens04}. More important for
QKD, the problem of the minimization of expectation values of
witness operators $W$ with respect to pure product states,
\begin{equation}\label{opt2}
\min_{\ket{a,b}}\ Tr(W\ \ketbra{a,b}),
\end{equation}
can also be formulated as a convex optimization problem
\cite{jens04}. Solving Eq.~(\ref{opt2}) allows to obtain new
witness operators which are finer than $W$, even within a
restricted class $W_{\cal C}$ of them \cite{jens04}. To see this,
let $W\in{}W_{\cal C}$ and denote the result of the minimization
problem in Eq.~(\ref{opt2}) as $\epsilon=\min_{\ket{a,b}}
Tr[\ketbra{a,b}\ W]>0$. Then the unnormalized new witness operator
given by $\tilde{W}=W-\epsilon\openone$ is finer than $W$ and
moreover it is guarantee that $\tilde{W}\in{}W_{\cal C}$ since
the observable $\openone$ is always accessible. At first sight it
seems that this procedure requires to have already a valid
entanglement witness $W$ for the given QKD protocol. However,
this operator does not need to be an entanglement witness in the
strict sense, but can be also a positive operator from the
restricted set which is more easy to characterize than an
entanglement witnesses \cite{Full}. With respect to the
minimization problem itself, it has been shown that although the
polynomial constraints parametrizing the pure states $\ket{a,b}$
are non-convex and computationally expensive to handle, one can
apply results from relaxation theory of non-convex problems
\cite{Shor,Lasserre}, notably the method of Lasserre
\cite{Lasserre}, to find hierarchies of solutions to that problem
in a way that each step of the hierarchy is a better
approximation than the previous one \cite{jens04}. Each step
itself amounts to solving an efficiently implementable
semi-definite program \cite{Semi} and the hierarchy is
asymptotically complete, in the sense that the exact solution is
asympotically attained. This means that during several steps of
the optimization method, better witness operators $\tilde{W}$ can
be obtained from $W$, belonging to the same restricted class
$W_{\cal C}$.

Finally, let us mention that, of course, not all higher
dimensional QKD protocols require the use of NDEWs to derive a
necessary and sufficient condition for the presence of
entanglement in P(A,B). For instance, consider the class of EB
protocols in $H=\mathbb{C}^2\otimes{}\mathbb{C}^N$, where Alice
realizes projection measurements onto the eigenvectors of the two
Pauli operators $\sigma_x$ and $\sigma_z$. In all these EB
protocols the accessible witness operators satisfy the condition
$W=W^{T_A}$. One can show that this fact implies for the given
dimensionalities that it is a necessary condition for a state
$\rho$ to be detected by $W$ that the operator
$\Omega=\frac{1}{2}(\rho+\rho^{T_A})$ is a non-positive operator.
To prove this, note that $Tr(W \rho)=Tr(W \Omega)$. From the work
by Kraus et al.~\cite{Kraus00} we learn that whenever $\Omega$ is
a non-negative operator it represents a separable state. To
summarize this remark, we learn that all detectable states for
this class of protocols are negative partial transposed (NPT)
entangled states and can be detected by using {\it only} DEWs. The
situation changes once Alice performs also a projection
measurement onto the eigenvectors of $\sigma_y$.

\subsection{Outlook to practical QKD systems}

The idea to check for quantum correlations in the observed data
with the help of a verification set of witnesses applies also, in
principle,  to real implementations of QKD setups \cite{expQKD}.
One can incorporate any imperfection of the sources and the
detection devices into the corresponding investigation within the
framework of trusted devices. In that framework one characterizes
detection devices by the use of an appropriate POVM description,
e.g. on the infinite dimensional Hilbert space of optical modes.
For P\&M schemes, one has to characterize additionally the given
source via the reduced density matrix of the virtual internal
preparation state, as described before. This idea then needs to
be generalized to signal states that are described by mixed
quantum states. For this purpose, one uses still a pure state as
internal preparation, but Alice's signal preparation corresponds
now no longer to a projection onto an orthogonal set of pure
states, but to projections onto orthogonal subspaces, thereby
effectively preparing mixed states.

In those general scenarios, it will be difficult to provide the
minimal verification set of witnesses. Instead, one can fall back
to the approach to search for just one accessible witness via
numerical methods such as presented in the previous section. In
this way, one can search through restricted classes of accessible
witnesses at the price that the result of this search will not be
conclusive, i.e. this search yields only a sufficient condition.

\section{CONCLUSION}\label{Concl}

A necessary precondition for secure quantum key distribution
(QKD) is that sender and receiver can use their available
measurement results to prove the presence of entanglement in a
quantum state that is effectively distributed between them.
Moreover, this result applies both to {\it prepare\&measure}
(P\&M) and {\it entanglement based} (EB) schemes. This means that
to construct practical and efficient new QKD protocols, it is
vital to separate the generation of two-party correlations from
the public discussion protocol which extracts a key from those
data. Among all separability criteria to deliver this necessary
entanglement proof, entanglement witnesses (EWs) are specially
appropriate since from them one can derive a necessary and
sufficient condition for the existence of quantum correlations
even when the state shared by the users cannot be completely
reconstructed.

In a recent work \cite{Curty04}, the set of optimal witness
operators for two well-known EB schemes: the $6$-state and the
$4$-state EB protocols, was obtained and a necessary and
sufficient condition to detect entanglement in both protocols was
derived. The purpose of this paper was to complete these results,
now showing specifically the analysis for the case of P\&M
schemes where, contrary to the case of EB schemes, now the
reduced density matrix of the sender is fixed and cannot be
modified by the eavesdropper. In particular, we have investigated
the signal states and detection methods of the $4$-state and the
$2$-state P\&M schemes, and we have obtained a reduced set of EWs
that can be used to provide a necessary and sufficient condition
for the existence of quantum correlations in both protocols.

Finally, we have discussed very briefly how to detect quantum
correlations in higher dimensional QKD schemes and in practical
QKD, where the characterization of EWs is not as easy to handle
as in the case of qubit-based QKD schemes. In this scenario it
might be still of interest to obtain one relevant EW as a first
step towards the demonstration of successful QKD. Here,
mathematical results from the field of convex optimization theory
can be used to get new insights in the construction of finer EWs
within a given class.

\section{ACKNOWLEDGEMENTS}
The authors wish to thank A. Dolinska, Ph. Raynal, and especially
K. Tamaki for very useful discussions, P. van Loock for his
comments on the manuscript, and P. Horodecki for helpful
discussions about Section \ref{QC}. This work has been supported
by the DFG (Emmy Noether programme, and the Graduiertenkolleg
$282$), the European Commission (Integrated Project SECOQC), and
the network of competence QIP of the state of Bavaria.

\appendix

\section{Condition for $W_2\ngeq{}0$}\label{A}

In this appendix we provide a proof for Observation
\ref{positive}. Our starting point is the most general form of the
operators
$W_2=|0\rangle\langle{}0|\otimes{}A+|1\rangle\langle{}1|\otimes{}B+x\
C(\theta)$. That is, if we write the matrix element of $W_2$
explicitly we have \be W_2= \left(
\begin{array}{cccc}
a & c & xe^{i\theta} & 0 \\
c & b & 0 &xe^{i\theta} \\
xe^{-i\theta} & 0 & e & g \\
0 & xe^{-i\theta} & g & f
\end{array}
\right). \label{redcrit1} \ee

Now we have to understand what the condition $W_2 \geq 0$ imposes
on the elements of $W_2.$ We have by definition of the class
$W_2$: $A \geq 0$ and $B \geq 0$ or, since $A$ and $B$ are of
full rank, we can assume at this point $A
> 0$ and $B > 0.$ Under this assumption, we can use the well
known fact that such a block matrix as in Eq.~(\ref{redcrit1}) is
positive iff its Schur complement is positive
\cite{hornjohnson1}, which implies here \be A-x^2B^{-1}\geq 0.
\ee Introducing the notation $y=x^2/det(B)$ the $2\times 2$
matrix \be X= \left(
\begin{array}{cc}
a-yf& c+yg\\
c+yg & b-ye
\end{array}
\right) \geq 0 \ee has to be positive. This is the case, iff
$det(X)\geq 0$ and $Tr(X)\geq 0.$ After a short calculation, this
implies that \bea &&\!\!\!\!\!\! ab-c^2-(ae + bf + 2cg)y + (ef -
g^2) y^2 \geq 0, \label{redeq1}
\\
&&\!\!\!\!\!\! (a+b)-y(e+f)\geq 0. \label{redeq2} \eea When are
these inequalities fulfilled for different values of $y\geq 0$?
For $y=0$ the matrix $X$ is clearly positive. If $y$ increases
$X$ get first one negative eigenvalue. Thus Eq.~(\ref{redeq1}) is
violated, while Eq.~(\ref{redeq2}) is still valid. If $y$
increases further,  the eigenvalues decrease and
Eq.~(\ref{redeq2}) gets violated. Finally, $X$ gets two negative
eigenvalues, and Eq.~(\ref{redeq1}) is valid, while
Eq.~(\ref{redeq2}) is still violated. So we have to look for the
smallest zero of  Eq.~(\ref{redeq1}) which is given by \be
y_0=\frac{\alpha}{2 det(B)}- \sqrt{\frac{\alpha^2}{4
det(B)^2}-\frac{det(A)}{det(B)}}, \ee where we have used
$\alpha=Tr(AB)=(ae + bf + 2cg).$ We obtain, therefore, that $W_2$
is positive iff \be x \leq x_{min}, \ee with \be
x_{min}=\sqrt{\frac{\alpha}{2}-\sqrt{\frac{\alpha^2}{4}-{det(A)}{det(B)}}}.
\ee

\bibliographystyle{apsrev}

\begin{thebibliography}{99}
\bibitem{Aut} It is necessary to require an initial secret key to
authenticate the public communication channel between Alice and
Bob. They need to be sure that Eve is not trying to impersonate
them (``man-in-the-middle-attack'').
\bibitem{PartialTrans} Given an operator $O\in H_{AB}$ and an orthonormal
basis $\{|\alpha_i\rangle\}\in H_B$, with $i=1,...,N$, the partial
transposed of $O$ with respect to the subsystem $B$ in that basis
is defined as
\begin{equation}
O^{T_B}=\sum_{i,j=1}^N \langle{}\alpha_i|O|\alpha_j\rangle\
|\alpha_j\rangle\langle{}\alpha_i|
\end{equation}
In the same way, one can also define the partial transposed of
$O$ with respect to the subsystem $A$.
\end{thebibliography}

\end{document}